\documentclass[11pt,onecolumn,superscriptaddress,showpacs,nofootinbib,notitlepage]{article}
\usepackage{amsmath}
\usepackage{latexsym}
\usepackage{amssymb}
\usepackage{graphicx}
\usepackage[colorlinks=true, citecolor=blue, urlcolor=blue]{hyperref}
\usepackage{float}
\usepackage{amsfonts}
\usepackage{textcomp}
\usepackage{authblk}
\title{Triggers for cooperative behavior in the thermodynamic limit: a case study in Public goods game}
\author[1]{Colin Benjamin}
\author[1]{Shubhayan Sarkar}
\affil[1]{School of Physical Sciences, National Institute of Science Education and Research, HBNI, Jatni- 752050, India}
\begin{document}
\maketitle
\begin{abstract}
In this work, we aim to answer the question- what triggers cooperative behaviour in the thermodynamic limit by taking recourse to the Public goods game. Using the idea of mapping the 1D Ising model Hamiltonian with nearest neighbor coupling  to payoffs in game theory we calculate the Magnetisation of the game in the thermodynamic limit. We see a phase transition in the thermodynamic limit of the two player Public goods game. We observe that punishment acts as an external field for the two player Public goods game triggering cooperation or provide strategy, while cost can be a trigger for suppressing cooperation or free riding. Finally, reward also acts as a trigger for providing while the role of inverse temperature (fluctuations in choices) is to introduce randomness in strategic choices.
\\
{\bf Keywords:} Nash equilibrium; Public goods game; Ising model
\\
\end{abstract}

%\linenumbers
\section{Highlights}
{\bf In the context of evolution, it is observed that cooperation among individuals exists even when defection should be the choice for every player. In this work, we figure out what triggers this cooperative behavior in the thermodynamic limit by considering Public goods game both with and without punishment.

In recent years there have been some attempts to explain why individuals in a population cooperate even when defection would be a better choice. There has been a previous attempt by Adami and Hintze in Ref.~[5] to answer this question using the Ising model. In this work, we first point out the errors in their approach and then give the correct approach to solve this problem using an exact mapping to the 1D Ising model. We identify the parameters which trigger cooperative behavior among individuals for Public goods game both with and without punishment. We find that reward and punishment are the strongest triggers for promoting cooperative or provide behavior. However, in contrast, cost of the resource acts as a suppressor of cooperation or promotes defection, i.e., free riding. Inverse temperature (fluctuation in choices) introduces randomness in strategic choices.}
\section{Introduction}
Game theory aims to find an equilibrium strategy where both of the players are at the maximum benefit or the least loss. This is known as the Nash equilibrium \cite{5}. In games such as Prisoner's dilemma, defection is the Nash equilibrium for both the players. Under certain conditions like kin selection or reciprocal altruism \cite{11,12} cooperation becomes the  preferred choice in the Prisoner's dilemma game. However, in Hawk-Dove, frequently used to model cooperation among humans, and animals, the Nash equilibrium is for one-player to defect while the other to cooperate. In the real world, however we see that organisms do cooperate among each other like sharing of resources. Thus, in the thermodynamic limit cooperation is indeed a choice in the long run otherwise the population as a whole won't survive. An account for connections between evolution and game theory can be found in Ref. \cite{19}. In this paper we try to investigate what triggers cooperation in the thermodynamic limit of a generic two player game like Public goods game.    
\subsection{{Motivation}}
What happens in the thermodynamic limit for games like Prisoner's dilemma, Hawk-Dove, etc. is an outstanding problem of game theory, since it is only in this limit that games can mimic the large populations of humans or animals. In this context and using the Ising model, Ref.~\cite{1} makes an attempt to analytically approach the thermodynamic limit of the two player Prisoner's dilemma and three player Public goods game. However, we have shown in a previous work \cite{8} that this approach to the thermodynamic limit of games has some inconsistencies. In Ref.~\cite{8}, we extended the idea of mapping the 1D Ising model Hamiltonian to the payoffs in a game \cite{2} to rectify the mismatch between expected outcomes and the calculated results of Ref.~\cite{1}. In this paper we approach the thermodynamic limit using the method of Ref.~\cite{1} for the two player Public goods game with and without punishment and show the errors in the approach of Ref.~\cite{1}. Further, we analytically apporach the thermodynamic limit of the public goods game with and without punishment by using an exact mapping of the 1D Ising model to the public goods game.  We calculate the game magnetisation, i.e., the difference between fraction of population choosing a particular strategy say provide over free riding in the thermodynamic limit of a Public goods game and analyse it for triggers which lead to a phase transition or cooperative behaviour. We unravel a cost dependent phase transition along with triggers for cooperation such as reward and punishment. The most important difference between our approach and the traditional approach is that we tackle the problem analytically, i.e., we give a very simple analytical  formula for calculating the distribution of population depending on the strategies. The traditional methods, see Ref.~[3], are numerical and involve solving differential equations (replicator approach) but in our approach if the payoff matrix of the game is known then the equilibrium condition can be calculated directly. Further, the traditional methods are dynamical, i.e., involve time. Our approach is not dynamical. The main attraction of our work is not just what happens at the thermodynamic limit but that providers (or, cooperators) do emerge (a finite minority of providers/cooperators exist) at the thermodynamic limit which was previously believed not to happen and can also be inferred from Nowak's paper, see Ref.~[3].

This paper is organized as follows- first we review 1D Ising model and the analogy of Ref. \cite{2} for a general two player two strategy payoff matrix with the two spin Ising Hamiltonian. In section III, we deal with the two player Public goods game in the thermodynamic limit. We see a mismatch between expected outcomes and the calculated results when we approach the problem using the method of Ref. \cite{1}. This mismatch is resolved by employing the procedure as adopted in Ref. \cite{8} to tackle the thermodynamic limit of games. In section IV, we first apply the method of Ref. \cite{1} to calculate the Nash equilibrium for the Public goods  game with punishment in the thermodynamic limit. As before we again see a contradiction, which is resolved by taking recourse to the method of Ref. \cite{8} in the thermodynamic limit. We observe an additional feature that cost can also act as the external magnetic field in the same two player Public goods game in the thermodynamic limit, we end with conclusions.
\section{Connections between game theory and 1D Ising model}\label{sec2}
\begin{figure}[h!]
\begin{center}
\includegraphics[width=\linewidth]{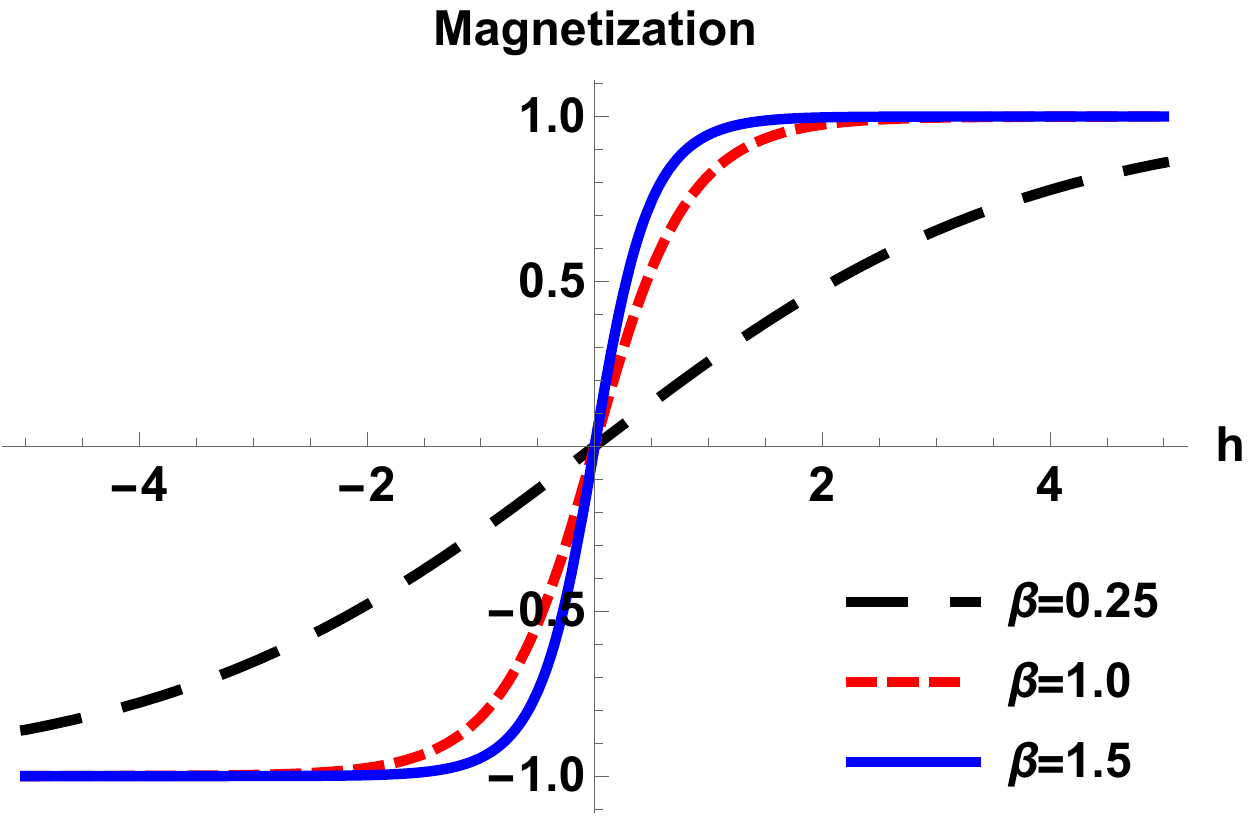}
  \caption{Magnetization vs. external magnetic field $h$ for 1D Ising model $(J=.1)$}
  \label{fig1:}
  \end{center}
\end{figure}

The Ising model \cite{6} consists of discrete variables that represent magnetic dipole moments of atomic spins that can be in one of two states $+1$ ($\uparrow$)  or $−1$ ($\downarrow$). In the 1D Ising model the spins at each site talk to their nearest neighbors only. The Hamiltonian of the 1D Ising model for $N$ sites is given as-
\begin{equation}\label{eq10}
H=-J\sum^N_{k=1}\sigma_k\sigma_{k+1}-h\sum^N_{k=1}\sigma_k.
\end{equation}
Herein, $h$ defines the external magnetic field while $J$ denotes the exchange interaction between spins the spins with the spins denoted as $\sigma$'s. The partition function $Z$, encodes the statistical properties of a system in thermodynamic equilibrium. For 1D Ising model, $Z$ is a function of temperature and parameters such as spins, coupling between spins and the external magnetic field. Thus the partition function can be defined as the statistical distribution of a system at thermal equilibrium and it follows Boltzmann statistics. The partition function for the 1D Ising model is then-
\begin{equation}\label{eq1}
Z=\sum_{\sigma_1}...\sum_{\sigma_N}e^{\beta(J\sum^N_{k=1}\sigma_i\sigma_{k+1}+h/2\sum^N_{k=1}(\sigma_k+\sigma_{k+1}))}
\end{equation}
$\beta=\frac{1}{k_B T}$. $\sigma_k$ denotes the spin of the $k^{th}$ site which can be either up (+1) or down (-1). The above sum is carried out by defining the transfer matrix $T$ with the following elements as-
\begin{eqnarray}
<\sigma|T|\sigma'>&=& e^{\beta(J\sigma\sigma'+\frac{h}{2}(\sigma+\sigma'))}\nonumber.
\end{eqnarray}
By using the completeness relation and transfer matrix, the partition function from Eq.~(\ref{eq1}) in the thermodynamic limit can be written as-
\begin{equation}
Z=e^{N\beta J}(\cosh(\beta h)\pm \sqrt{\sinh^2(\beta h)+e^{-4\beta J}})^N.
\end{equation}
The Free energy is then $F=-k_{B}T\ln Z$. Free energy allows us to study the state of equilibrium of a system. The state of equilibrium of a system corresponds to one that minimizes its free energy
$F$. The relation between the partition function $Z$ and the Free energy $F$  is $F=-k_{B}T \ln Z$ or $Z=e^{-\beta F}$. The net magnetization is found by averaging the total magnetization over all the allowed energy levels which is same as the partial derivative of Free energy with respect to the external magnetic field. Magnetization for the 1D Ising model is then-
\begin{equation}\label{eq8}
M=-\frac{df}{dh}=\frac{\sinh(\beta h)}{\sqrt{\sinh^2(\beta h)+e^{-4\beta J}}}.
\end{equation}
A plot of the Magnetization vs. external magnetic field $h$ is shown in Fig. \ref{fig1:} for different $\beta$, the inverse temperature.
\\In Ref.~\cite{2}, it has been shown that the 1D Ising model Hamiltonian for two spins and the two player payoff matrix for a particular game have a one-to-one correspondence. We first understand the method of Ref. \cite{2} by taking a general two player payoff matrix as-
\begin{equation}\label{eq9-1}
U=\left(\begin{array}{c|cc} & s_1 & s_2 \\\hline s_1 & x,x' & y,y' \\  s_2 & z,z' & w,w'\end{array}\right),
\end{equation} 
where $U(s_i,s_j)$ is the payoff function with $x, y, z, w$ as the payoffs for row player and $x', y', z', w'$ are the payoffs for column player, $s_1$ and $s_2$ denote the choices available to the players. In our analysis we consider symmetric games where $x=x'$, $y=z'$, $z=y'$ and $w=w'$. Thus, knowing payoff of the row player, the column player's payoff can be inferred. In Ref. \cite{2} A transformation is made via addition of a factor $\lambda$ to the $s_1$ column and $\mu$ to the $s_2$ column. Thus,   
\begin{equation}\label{eq2}
U=\left(\begin{array}{c|cc} & s_1 & s_2 \\\hline s_1 & x+\lambda & y+\mu \\  s_2 & z+\lambda & w+\mu\end{array}\right).
\end{equation} 
As shown in Appendix 7.3, the Nash equilibrium of the game, Eq.~(\ref{eq9-1}) remains unchanged under such transformations. Following Ref.~\cite{2} and choosing the transformations as $\lambda=-\frac{x+z}{2}$ and $\mu=-\frac{y+w}{2}$.  The transformed matrix becomes
\begin{equation}\label{eq12}
U=\left(\begin{array}{c|cc} & s_1 & s_2 \\\hline s_1 & \frac{x-z}{2} & \frac{y-w}{2} \\  s_2 & \frac{z-x}{2} & \frac{w-y}{2}\end{array}\right).
\end{equation} 
Since our aim is to find the Nash equilibrium in the thermodynamic limit, we start by identifying the connection between the 1D Ising model Hamiltonian for two spins with the transformed payoff matrix as in Eq.~(\ref{eq12}). The Hamiltonian in Eq.~(\ref{eq10}) for $N=2$ is given as-
\begin{eqnarray}\label{eq21}
H&=&-J(\sigma_1\sigma_{2}+\sigma_2\sigma_{1})-h(\sigma_1+\sigma_{2}).\nonumber\\
&&\mbox{ The individual energies of the two spins can be inferred as-}\nonumber\\
E_{1}&=&-J\sigma_1\sigma_{2}-h\sigma_1\nonumber\\
E_{2}&=&-J\sigma_2\sigma_{1}-h\sigma_{2}.
\end{eqnarray}
In Ising model, the equilibrium condition implies that the energies of spins are minimized. For symmetric coupling as in Eq. (\ref{eq1}), the Hamiltonian $H$ is minimized with respect to spins $\sigma_1, \sigma_2$. This is equivalent to maximizing $-H$ with respect to $\sigma_1, \sigma_2$. Game theory aims to search for an equilibrium strategy(Nash equilibrium) which can be achieved by maximizing the payoff function $U(s_i,s_j)$ Eqs.~(\ref{eq9-1}-\ref{eq12}) with respect to the choices $s_i,s_j$. For the two player Ising model case this is same as maximizing $-E_i$, in Eq.~(\ref{eq21}) with respect to spins $\sigma_i,\sigma_j$. Thus, the Ising game matrix for the row player is (see Ref.~\cite{2} for derivation of Eq.~(\ref{eq13-})) is-
\begin{equation}\label{eq13-}
U_{Ising}=\left(\begin{array}{c|cc}  & s_2=+1 & s_2=-1  \\\hline s_1=+1 & J+h & -J+h \\s_1=-1 & -J-h & J-h\end{array}\right).
\end{equation}
We compare the Ising game matrix Eq.~(\ref{eq13-}) to the matrix elements of Eq.~(\ref{eq12}) to determine the values of $J$ and $h$. Thus,
\begin{eqnarray*}
J=\frac{x-z+w-y}{4},\ h=\frac{x-z+y-w}{4}, \mbox{and }
\end{eqnarray*} 
therefore  the game Magnetization from Eq.~(\ref{eq8}) can be written in terms of the payoff matrix elements Eq.~(\ref{eq9-1}) as-
\begin{equation}\label{eq13}
M=\frac{\sinh(\beta \frac{x-z+y-w}{4})}{\sqrt{\sinh^2(\beta \frac{x-z+y-w}{4})+e^{-\beta (x-z+w-y)}}}.
\end{equation} 
which is identified as "game magnetization`` in the thermodynamic limit of the game.

In Ising model, magnetization refers to the difference between number of spins pointing up versus the number pointing down. For  games we define a quantity akin to the magnetization called game magnetization(10) which refers to the difference between fraction of players opting for a particular strategy, say, Provide versus the fraction opting for Free ride in case of Public goods game. For the Ising model, $J$-the exchange coupling is in Joules, while $h$-applied magnetic field is also in Joules while $\beta=\frac{1}{k_{B}T}$ is in units of Joule$^{-1}$. In the case of games, payoffs are unit less and $\beta$ is also unit less and acts as a randomizing parameter.
It should be noted that $\beta$ is the inverse temperature ($1/k_BT$) in Ising model. Decreasing $\beta$ or increasing temperature randomizes or makes the spins more disordered. As discussed above, two player games have an analogy with two spin Ising Hamiltonian where spins are analogous to strategies. Thus, decreasing $\beta$ randomizes the strategic choices available to each player which effectively implies $\beta$ may trigger either cooperation or defection depending on the other parameters of the problem. This completes the connection of 1D Ising model to a general two player game. In the following sections we will apply this to the Public goods game both with and without punishment and analyze them in the thermodynamic limit.
\section{Public goods game}
The Public goods game otherwise known as the "free rider problem" is a social dilemma game akin to the Prisoner's dilemma game. A "public good" is a perfectly shareable resource, which once produced can be utilized by all in a community. In the two player version of the Public goods game, this "public good" can be produced by either player alone by paying the full cost of the service or it can be jointly produced if each pay for half of the service. The payoffs for the cooperators (provider) and defectors (free rider)\cite{7} are given by-
\begin{eqnarray}
P_D=kn_cc/N,\ \ P_C=P_D-c
\end{eqnarray}
where $c$ is the cost of the service, $k$ denotes the multiplication factor of the "public good", $N$ denotes the number of players in the group (in the two player Public goods game, N=2), and $n_c$ denotes the number of cooperators in the group.  Thus, the payoff matrix for the two player Public goods game can be written (for the case when both provide- $P_c=kc-c=2r$ and when both free ride-$P_D=0$ while when one free rides and another provides then is $kc/2=r+c/2$ for free rider and $kc/2-c=r-c/2$ for the provider):
\begin{equation}\label{eq26}
U=\left(\begin{array}{c|cc} & provide & free\ ride \\\hline provide & 2r,2r & r-\frac{c}{2},r+\frac{c}{2} \\  free\ ride & r+\frac{c}{2},r-\frac{c}{2} & 0,0\end{array}\right)
\end{equation}
where $r>0$ and $c>0$. As we can see from the payoff matrix, $(free ride,free ride)$ is the stable strategy or the Nash equilibrium for $r<c/2$. However, when $r>c/2$ $(provide,provide)$ is the Nash equilibrium. We first calculate the game magnetization in the thermodynamic limit for this two player public goods game using the approach of Ref.~\cite{1}, bringing out the imperfections in the approach of Ref.~\cite{1} and then do it correctly using our approach as elucidated in section 4.2.
\subsection{Connecting Ising model with Public goods game- approach of Ref.~\cite{1}}    
Now, we extend the approach of Ref.~\cite{1} to the two player public goods game without punishment. As described above, in the Public goods game the provide strategy can be written as cooperation and free ride strategy as defection. Thus, provide strategy or cooperation is represented as spin up $(\uparrow)$ and free ride strategy or defect as spin down $(\downarrow)$, which are represented as vectors, i.e., kets $|C>, |D>$ in bra-ket notation as- 
\begin{equation}
|C>=\left(\begin{array}{c} 1 \\ 0 \end{array}\right),|D>=\left(\begin{array}{c} 0 \\ 1\end{array}\right).
\end{equation}
 In matrix representation ket vectors are- \[|0\rangle = \left(\begin{array}{c} 1 \\ 0\end{array}\right) \mbox{and }  |1\rangle = \left(\begin{array}{c} 0 \\ 1\end{array}\right),\] while bra vectors \[ \langle 0|= \left(\begin{array}{cc}1 & 0\end{array}\right)^{T} \mbox{and } \langle 1|= \left(\begin{array}{cc}0 & 1\end{array}\right)^{T}, \mbox{T being transpose}.\]

Similar to Ising model, the Hamiltonian of the system can be written using the payoff matrix U, and the projectors $P^{i}_{0}=P^C=|0><0|$, and $P^{i}_{1}=P^D=|1><1|$. The projectors are defined as outer products of the bra-ket vectors as-  \[|0\rangle\langle 0|=\left(\begin{array}{c} 1 \\ 0\end{array}\right)\left(\begin{array}{cc} 1 & 0 \end{array}\right)=\left(\begin{array}{cc} 1 & 0\\ 0 & 0\end{array}\right), 
  |1\rangle\langle 1|=\left(\begin{array}{c} 0 \\ 1\end{array}\right)\left(\begin{array}{cc} 0 & 1 \end{array}\right)=\left(\begin{array}{cc} 0 & 0\\ 0 & 1\end{array}\right).\]

The Hamiltonian is then given by-
\begin{equation}
H=\sum_{N}^{i=1}\sum_{m,n=0,1}U_{mn}P_m^{(i)}\otimes P_n^{(i+1)},
\end{equation}
where $0$, and $1$ denote spin-up $(\uparrow)$, and spin-down $(\downarrow)$ sites and elements of payoff matris $U$ are:$U_{00}=2r$, $U_{01}=r-c/2$, $U_{10}=r+c/2$ and $U_{11}=0$. $N$ denotes the total number of players. 
The Kronecker product of two matrices $M_{1}=\left(\begin{array}{cc} A & B\\ C & D\end{array}\right)$ and $M_{2}=\left(\begin{array}{cc} A & B\\ C & D\end{array}\right)$ is $K=M_{1}\otimes M_{2}$, and is given by- \[K=\left(\begin{array}{cc} A & B\\ C & D\end{array}\right) \otimes \left(\begin{array}{cc} R & S\\ T & P\end{array}\right)\]
or, \[K=\left(\begin{array}{cc} A\left(\begin{array}{cc} R & S\\ T & P\end{array}\right) & B\left(\begin{array}{cc} R & S\\ T & P\end{array}\right)\\ C \left(\begin{array}{cc} R & S\\ T & P\end{array}\right) & D\left(\begin{array}{cc} R & S\\ T & P\end{array}\right)\end{array}\right)\!=\!\left(\begin{array}{cccc} AR & AS & BR & BS \\ AT & AP & BT & BP \\ CR & CS & DR & DS \\ CT & CP & DT & DP \end{array}\right).\]
We get the game magnetization in the thermodynamic limit using the approach of Ref.~\cite{1} to be-
\begin{equation}\label{eq4}
M = \frac{e^{-2\beta r}-1}{(1+e^{-2\beta r})}=-\tanh(\beta r).
\end{equation}
For a detailed calculation, of the game magnetization using the approach of Ref.~\cite{1}, see Appendix 7.1.
\begin{figure}[h!]
\begin{center}
\includegraphics[width=\linewidth]{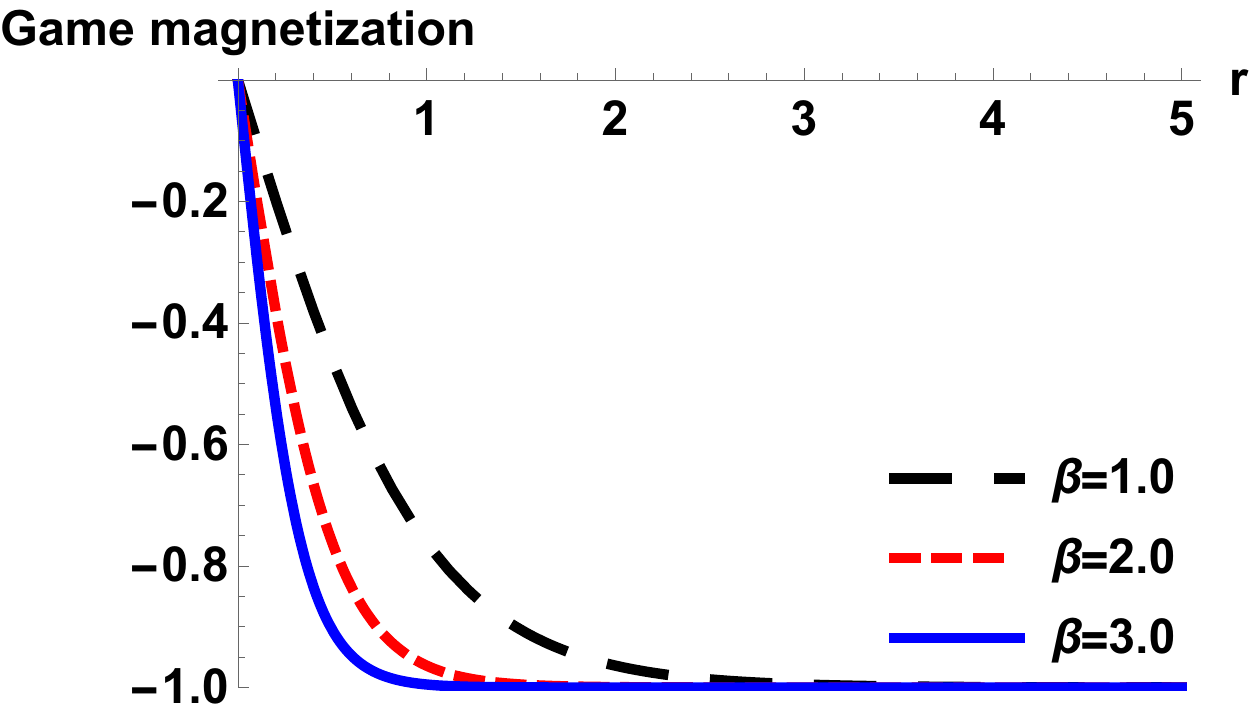}
  \caption{Game magnetization  vs. reward $r$ for the Public goods game for different values of $\beta$ using the method of Ref. \cite{1}. Note that the game magnetization is independent of $c$. Taking any value of c say $c=2$, we see that for $r>c/2$, the game magnetization is still negative which implies that free riding is the Nash equilibrium which obviously contradicts the solution for the two player Public goods game. Further when $r=0$, the  game magnetization is $0$ which again is a contradiction.}
  \label{fig7:}
  \end{center}
\end{figure}
As we can see from Fig.~\ref{fig7:}, the game magnetization using approach of Ref.~\cite{1} is always negative for $r>0$ which means that free ride or defect is the Nash equilibrium in the thermodynamic limit. However, from the payoff matrix Eq.~(\ref{eq26}), when $r$ is less than $c/2$ then free ride is the dominant strategy. However, when $r$ is greater than $c/2$, then provide or cooperation is the Nash equilibrium. So its expected that a phase transition should occur at $c/2$. Further, when $r=0$ from the payoff matrix Eq.~(\ref{eq26}), we see that the Nash equilibrium is still defect. However, with the approach of Ref.~\cite{1} we see that at $r=0$ the game magnetization is also 0 meaning that there are equal number of providers and free riders which is again a contradiction. In the next section, we show that using our approach to the problem the issues with the method of Ref.~\cite{1} are resolved.
\subsection{Connecting Ising model with Public goods game-the correct approach}
Following the calculations in section 3 and using the method of Ref.~\cite{8}, we make the correct connection of the Public goods game payoff matrix, Eq.~(\ref{eq26}) and Ising game matrix Eq.~(\ref{eq13-}). As in Eq.~(\ref{eq2}), we add $\lambda=-\frac{x+z}{2}=-\frac{6r+c}{4}$ to column 1 and $\mu=-\frac{y+w}{2}=-\frac{2r-c}{4}$ to column 2 of the payoff matrix, Eq.~(\ref{eq26}). Thus the Public goods game payoff matrix Eq.~(\ref{eq26}) reduces to-
\begin{equation}\label{trans-pg-payoff}
U=\left(\begin{array}{c|cc} & provide & free\ ride \\\hline provide & \frac{2r-c}{4} & \frac{2r-c}{4} \\  free\ ride & -\frac{2r-c}{4} & -\frac{2r-c}{4}\end{array}\right).
\end{equation} 
Comparing this to the Ising game matrix Eq.~(\ref{eq13-}), we have-
$J+h=\frac{2r-c}{4}$ and $J-h=-\frac{2r-c}{4}$. Solving these simultaneous equations, we get-
$J=0$ and $h=\frac{2r-c}{4}$. The game magnetization in the thermodynamic limit is then-
\begin{equation}
M=\frac{\sinh(\beta h)}{\sqrt{\sinh^2(\beta h)+e^{-4\beta J}}}=\tanh(\beta \frac{2r-c}{4}).
\end{equation} 
\begin{figure}[h!]
\begin{center}
\includegraphics[width=\linewidth]{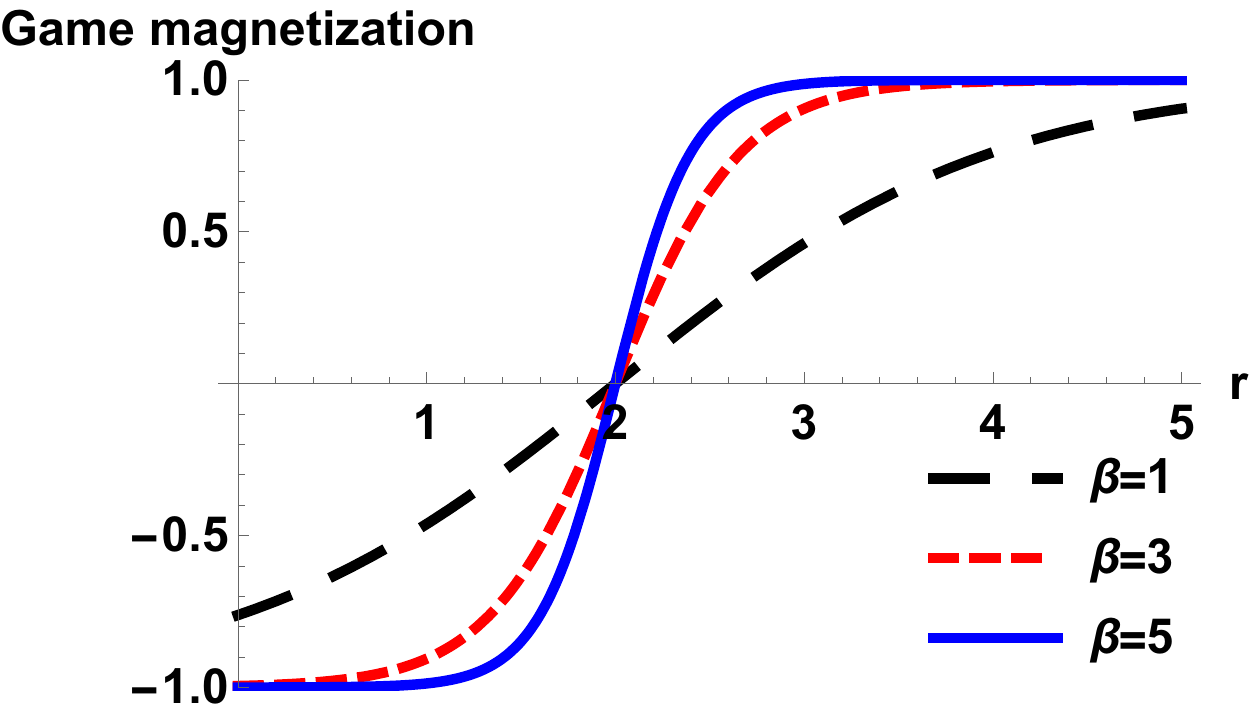}
  \caption{Game magnetization vs. reward $r$ for Public goods game for different values of $\beta$ and cost $c=4$. For same cost of ``public good" $(c)$, the critical point doesn't change.}
  \label{fig3:}
  \end{center}
\end{figure}
\begin{figure}[h!]
\begin{center}
\includegraphics[width=\linewidth]{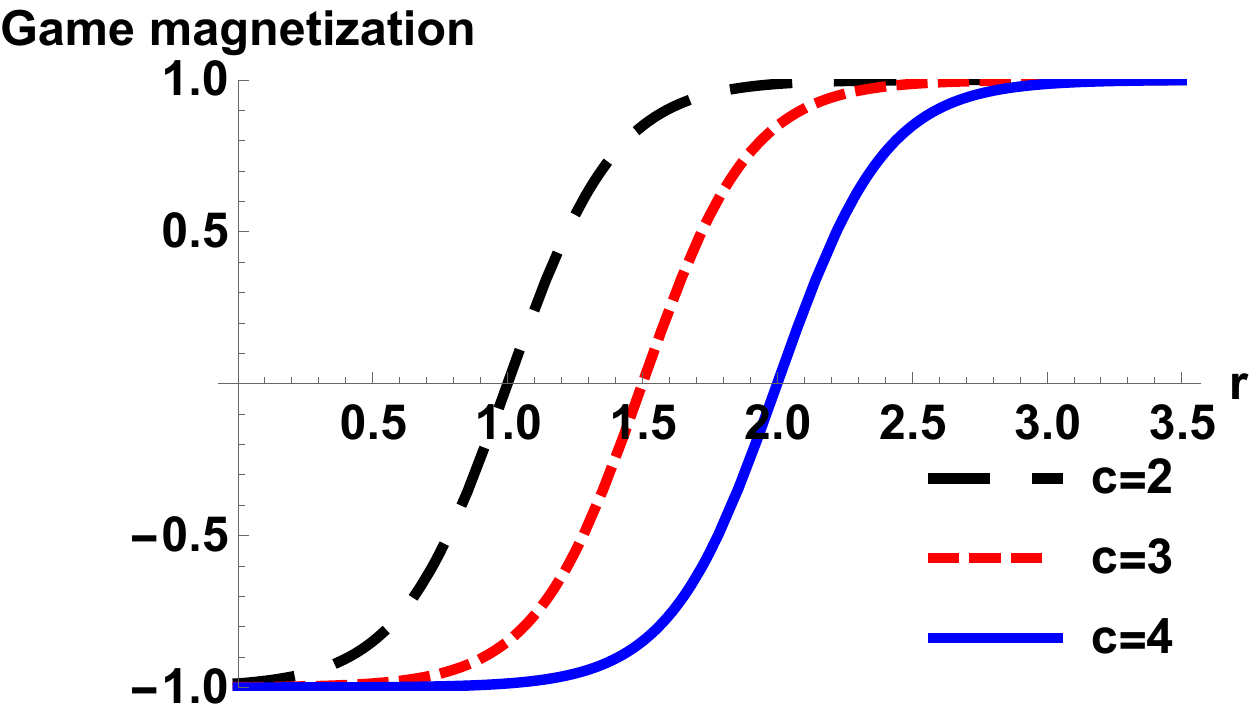}
  \caption{Game magnetization vs. reward $r$ for Public goods game for different values of the cost $c$ with $\beta=5$. As the cost of ``public good" $(c)$ increases, for the same reward more players choose to free ride or defect.}
  \label{fig4:}
  \end{center}
\end{figure}
As we see in Fig.~\ref{fig3:}, there is a phase transition which occurs at $r=c/2=2$ for $c=4$. For $r<c/2$ the game magnetization is negative, i.e., defection or free ride is the Nash equilibrium and for $r>c/2$, game magnetization is positive, i.e., the Nash equilibrium is provide strategy. $\beta$ which defines the randomness in the strategic choices available to the players has no bearing on the critical point of the phase transition. As $\beta\rightarrow 0$ the game magnetization tends to vanish, i.e., the number of cooperators and defectors become identical. Further, in the regime $r<2$, as we decrease $\beta$ from 5 to 1, almost one-fourth of the population start cooperating. In contrast to $\beta$, the cost $(c)$ as in Fig.~\ref{fig4:} has a bearing on the critical point of the phase transition. Herein, we see increasing cost for a fixed $\beta$ ($\beta=5$) propels the critical point to higher values of reward $(r)$. This is because when cost of ``public good" increases with the reward remaining constant fewer players will pay the cost. Further, when reward $(r)$ increases keeping the cost $(c)$ constant, the number of cooperators increases. 

For two player games, both the players would choose the Nash equilibrium strategy. A natural extension from two player case to $N$ players would be that all the players would go for the Nash equilibrium strategy. However, in contrast to the two player case this is not what we observe. Its true that in the thermodynamic limit the Nash equilibrium strategy is chosen by majority of the players  but there are exceptions. For example, in Public goods game when reward increases but is less than half of the cost of ``public good", even when the Nash equilibrium strategy is defection but still the number of cooperators increases. There is always a small fraction of players who choose to cooperate in the thermodynamic limit and that fraction increases as the reward $r$ increases. Further, we see that for the case when cost becomes very high, there are individuals in the population who pay for the ``public good" even when defection would be the best choice. The game magnetization we calculate is defined as the net difference in the fraction of players opting for a particular strategy, say Provide in Public Goods game versus the fraction opting for free riding. Both these strategic choices are akin to phases of the Ising model. A phase transition occurs in the game when majority of population opts to change their particular strategy, as at $r=c/2$. For $r<c/2$, majority free ride while for $r>c/2$ the majority provide.
\section{Public goods game with punishment}
In the Public goods game, the punishment $p$ is introduced such that whenever a player defects or free rides he has an additional negative payoff given by $-p$. Thus, the modified payoff matrix, from Eq.~(\ref{eq26}), is-
\begin{equation}\label{eq27}
U=\left(\begin{array}{c|cc} & provide & free\ ride \\\hline provide & 2r,2r & r-\frac{c}{2},r+\frac{c}{2}-p \\  free\ ride & r+\frac{c}{2}-p,r-\frac{c}{2} & -p,-p\end{array}\right),
\end{equation}
where $r,c,p>0$. As we can see from the payoff matrix Eq. (\ref{eq27}), when $r>c/2-p$, then cooperation or provide is the Nash equilibrium, but when $r<c/2-p$ then defection or free riding is the Nash equilibrium. We first calculate the game magnetization in the thermodynamic limit for this two player Public goods game with punishment using the approach of Ref. \cite{1}, and point out the imperfections and then do it correctly using our approach as alluded to in sections 3 and 4.2.
\subsection{Connecting Ising model and Public goods game with punishment using approach of Ref.~\cite{1}}    
\begin{figure}[h!]
\begin{center}
\includegraphics[width=\linewidth]{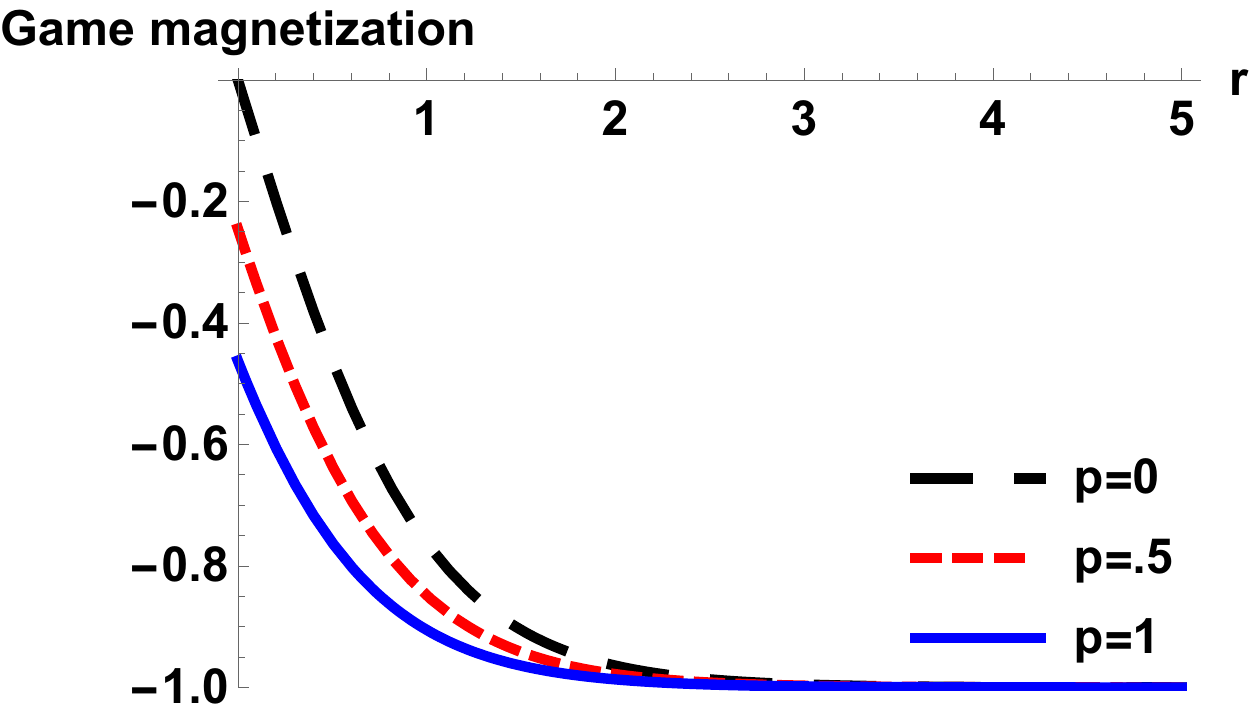}
  \caption{Game magnetization vs. reward $r$ for the Public goods game with punishment for different values of the cost $p$ with $\beta=1$ using the method of Ref.~\cite{1}. We see that there is no phase transition. Also when increasing the punishment $p$, the  game magnetization becomes more negative which is incorrect.}
  \label{fig8:}
  \end{center}
\end{figure}
As described earlier, in the Public goods game provide can be written as cooperation and free ride as defection. Similar to 1D Ising model, the Hamiltonian of the system using the payoff matrix $U$, Eq.~(\ref{eq27}) and the projectors $P^{i}_{0}=P^C=|0><0|$ and $P^{i}_{1}=P^D=|1><1|$ is given by-
\begin{equation}
H=\sum_{N}^{i=1}\sum_{m,n=0,1}U_{mn}P_m^{(i)}\otimes P_n^{(i+1)},
\end{equation}
where 0 denotes spin up $(\uparrow)$ and 1 denotes spin down $(\downarrow)$ sites with payoff matrix elements: $U_{00}=2r$, $U_{01}=r-c/2$, $U_{10}=r+c/2-p$ and $U_{11}=-p$. Similar to that shown in section 4.1, we find the game magnetization in the thermodynamic limit using the approach of Ref.~\cite{1} to be-
\begin{equation}\label{eq41}
M = \frac{e^{-2\beta r}-e^{\beta p}}{(e^{\beta p}+e^{-2\beta r})}=-\tanh(\beta (r+p/2)).
\end{equation}
As we can see from Fig.~\ref{fig8:}, the game magnetization is always negative for $r>0$ and $p>0$, which means that free ride is always the Nash equilibrium in the thermodynamic limit. Further, when the punishment increases higher fraction of players chooses to defect which is an incorrect conclusion. However, when we analyze the situation, from the payoff matrix Eq.~(\ref{eq41}) if the punishment $p$ is such that reward $r>c/2-p$ then the players should choose provide. Further, the game magnetization is independent of the cost which is unexpected. In the next section we resolve these issues and find the correct game magnetization.
\subsection{Connecting Ising model and Public goods game with punishment- the correct approach}
Following the calculations in sections 3 and 4.2 and using the method of Ref.~\cite{8}, we make the correct connection of Public goods game with punishment payoff matrix Eq.~(\ref{eq27}) and Ising game matrix Eq.~(\ref{eq13-}). In our approach as elucidated in Eq.~(\ref{eq2}), we add  $\lambda=-\frac{x+z}{2}=-\frac{6r+c-2p}{4}$ to column 1 and $\mu=-\frac{y+w}{2}=-\frac{2r-c-2p}{4}$ to column 2 of the payoff matrix Eq.~(\ref{eq27}) to make the mapping between the payoff matrix of game and 2-spin Ising game matrix, Eq.~(\ref{eq13-}) exact. Thus, the transformed Public goods game payoff matrix Eq.~(\ref{eq27}) reduces to-
\begin{equation}
U=\left(\begin{array}{c|cc} & provide & free\ ride \\\hline provide & \frac{2r-c+2p}{4} & \frac{2r-c+2p}{4} \\  free\ ride & -\frac{2r-c+2p}{4} & -\frac{2r-c+2p}{4}\end{array}\right)
\end{equation} 
Comparing this to the Ising game matrix- Eq.~(\ref{eq13-}), we have $J+h=\frac{2r-c+2p}{4}$, and
$J-h=-\frac{2r-c+2p}{4}$. Solving these simultaneous equations we get- $J=0$ and $h=\frac{2r-c+2p}{4}$.
Thus, the game magnetization for Public goods game with punishment, from Eq.~(\ref{eq13}) is-   
\begin{equation}\label{eq23}
M=\frac{\sinh(\beta h)}{\sqrt{\sinh^2(\beta h)+e^{-4\beta J}}}=\tanh(\beta \frac{2r-c+2p}{4})
\end{equation} 
\begin{figure}[t]
\begin{center}
\includegraphics[width=\linewidth]{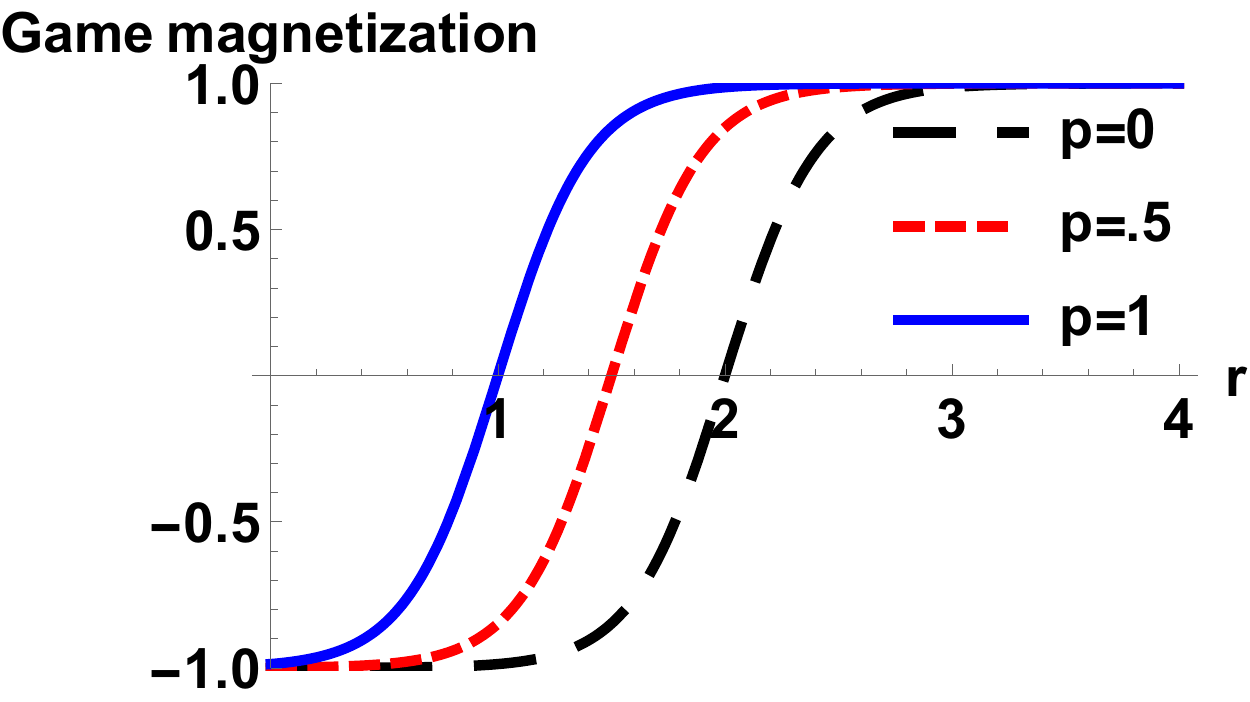}
  \caption{Game magnetization vs. the reward $r$ for classical Public goods game with punishment for different values of the punishment $p$, cost $c=4$ with $\beta=5$. As the punishment increases, for same value of reward more players choose to provide or cooperate.}
  \label{fig5:}
  \end{center}
\end{figure}
In Fig.~\ref{fig5:} we plot game magnetization Eq.~(\ref{eq23}) versus the reward $r$ for different values of punishment. We see that as the punishment $p$ increases, considering $\beta$ fixed ($\beta=5$), the critical point for the phase transition decreases to a lower value of reward $(r)$. This is because when punishment for defecting increases, with the reward remaining constant, more players would provide, as the penalty for defecting is high. In contrast to the cost of the ``public good", increasing punishment for defection increases the number of cooperators.

Similar to the Public goods game without punishment, we see that not all players go for the Nash equilibrium strategy in the infinite player or thermodynamic limit of Public goods game with punishment. For example, when the punishment is low as compared to the reward, then higher fraction of players choose to free ride or defect but there exist a finite fraction of players who cooperate or provide, even when punishment $p$ increases from $0$ to $1$, the fraction of cooperators at $r=1$ increases by almost 50 $\%$. $\beta$ has no role in determining the critical point in the phase transition although depending on other triggers like cost, reward or punishment changing $\beta$ may lead to an increase of cooperators in certain situations.
\section{Conclusions}
We have studied the connection between Ising model and game theory to find the triggers for cooperative behavior in the Public goods game with and without punishment. We contrasted the results from our approach with that of Adami, Hintze's in Ref.~\cite{1}. We unravel some inconsistencies in the method of Ref.~\cite{1}. In the Appendix sections 7.1 and 7.2,  a detailed discussion on the errors in Adami-Hintze's method\cite{1} is presented. Further, using the correct approach to the problem as dealt with in sections 3, 4.2 and 5.2, we see that in the Public goods game cost plays a non trivial role in determining the critical point of the phase transition between providing and free riding. The thermodynamic limit of the Public goods game both with and without punishment shows that reward and punishment are the strongest triggers for providing. Cost invariably suppresses cooperative or provide  while the role of $\beta$ (fluctuation in choices) is to randomize strategic choices. 
\section{Appendix}
\subsection{ Derivation of game magnetization for the Public goods game without punishment using the method of Ref.~\cite{1}}
Following the method of Ref.~\cite{1}, the Hamiltonian is given by-
\begin{equation}
H=\sum_{N}^{i=1}\sum_{m,n=0,1}U_{mn}P_m^{(i)}\otimes P_n^{(i+1)}.
\end{equation}
where $U_{mn}$ denotes the matrix elements of the payoff matrix Eq.~(\ref{eq26}). Thus, the partition function is
\begin{equation}
Z=\sum_{x}\langle x|e^{-\beta H}|x\rangle=\sum_{m_1,m_2...m_n}e^{-\beta(U_{m_1m_2}+U_{m_2m_3}...+U_{m_nm_1})}= Tr(E^N)\nonumber,
\end{equation}
where $|x\rangle=|m_1m_2....m_n\rangle$ is the state of the system and the $ij^{th}$ element of the matrix $E$ is $e^{-\beta U_{ij}}$, with the $E$ matrix given as-
\begin{equation}\label{eqp2}
E=\left(\begin{array}{cc} e^{-2\beta r} &  e^{\beta (r-\frac{c}{2})} \\e^{-\beta (r+\frac{c}{2})} & 1\end{array}\right).
\end{equation}
Using the above expression, the partition function becomes-
\begin{equation}
Z=Tr(e^{-\beta H})=Tr(E^{N})=(1+e^{-2\beta r})^N.
\end{equation}
As in Ref.~\cite{1}, the average value of choosing a particular strategy $m$ is the expectation value of $P^i_{m}$ with $m$ being the spin at site $i$. Thus,
\begin{equation}
\langle P^i_{m}\rangle=\frac{\sum_{x}\langle x|P^i_{m}e^{-\beta H}|x\rangle}{ZN}.
\end{equation}
The average value $\langle P^{C}\rangle=\langle P^i_{0}\rangle$ for spin up $(\uparrow)$, i.e., $m=0$ or provide strategy(C) is-
\begin{equation}
\langle P^C\rangle=\frac{\sum_{x}<x|P^Ce^{-\beta H}|x>}{ZN}=N\frac{Tr(E^CE^{N-1})}{ZN}\nonumber,
\end{equation}
where 
\begin{equation}
E^C=\left(\begin{array}{cc} e^{-\beta r} & e^{-\beta (r-\frac{c}{2})}\\  0& 0\end{array}\right).
\end{equation}
Thus, $Tr(E^CE^{N-1})=Tr(E^C)Tr(E^{N-1})=e^{-2\beta r }(1+e^{-2\beta r })^{N-1}$, implying-
\begin{equation}
\langle P^C \rangle=\frac{e^{-2\beta r }}{(1+e^{-2\beta r })}\nonumber.
\end{equation}
Similarly, we can calculate the average value $\langle P^{D}\rangle=\langle P^i_{1}\rangle$ for spin down $(\downarrow)$, i.e., $m=1$ or free ride strategy(D)-
\begin{equation}
\langle P^D\rangle=\frac{\sum_{x}\langle x|P^D e^{-\beta H}|x\rangle}{Z}=N\frac{Tr(E^DE^{N-1})}{ZN}\nonumber,
\end{equation}
where
\begin{equation}
E^D=\left(\begin{array}{cc} 0& 0 \\ e^{-\beta (r+\frac{c}{2})}& 1 \end{array}\right).
\end{equation}
Thus, $Tr(E^DE^{N-1})=Tr(E^D)Tr(E^{N-1})=(1+e^{-2\beta r })^{N-1}$, implying-
\begin{equation}
\langle P^D \rangle=\frac{1}{(1+e^{-2\beta r})}\nonumber.
\end{equation}
The game magnetization $M$, i.e., the difference in the average fraction of players choosing cooperation over defection using approach of Ref.~\cite{1} is then-
\begin{equation}
M=\frac{e^{-2\beta r}-1}{(1+e^{-2\beta r})}=-\tanh(\beta r).
\end{equation}
A similar calculation can be done for the two player Public goods game with punishment where we use the payoff matrix $U$ as in Eq.~(\ref{eq27}).
\subsection{The error in Adami-Hintze's approach, Ref.~\cite{1}-}
Herein we first analyze the two player analog of Adami-Hintze's approach in order to expose the mistakes in their method.\\ 
{\bf{Adami and Hintze's approach}}\\ 
We start with a two spin system and then calculate the average payoff. The Hamiltonian for a 2-spin system defined using the Adami-Hintze approach is- 
\begin{equation}
H=\sum_{m,n=1,2}U_{mn}P_m^{(1)}\otimes P_n^{(2)}+\sum_{m,n=1,2}U_{mn}P_m^{(2)}\otimes P_n^{(1)}
\end{equation}
Equation(30) is Eq.~(2) of Ref.~[5] for $N=2$. $U_{mn}$ represents the elements of the payoff matrix $U$ for the $m^{th}$ row and $n^{th}$ column. The operator $P^{1}_{m}$ defines the projector of site $1$ with $m$ being $1$ (meaning  $P_1^{1}= |0\rangle\langle0|$) or if at site $1$ again with $m=2$ then ( $P_2^{1}=|1\rangle\langle1|$). Similarly $P_m^{2}$  signifies the projector of site $2$ with $m$ taking values $1$ or $2$.  $m, n$ denote  the indices for strategies which for $1$ denotes  cooperation while for $2$ denotes defection. Expanding the Hamiltonian Eq.~(30)-
\begin{eqnarray}
H=U_{11}P_1^{(1)}\otimes P_1^{(2)}+ U_{12}P_1^{(1)}\otimes P_2^{(2)}+
U_{21}P_2^{(1)}\otimes P_1^{(2)}+
U_{22}P_2^{(1)}\otimes P_2^{(2)}\nonumber\\ +U_{11}P_1^{(2)}\otimes P_1^{(1)}+ U_{12}P_1^{(2)}\otimes P_2^{(1)}+ 
U_{21}P_2^{(2)}\otimes P_1^{(1)}+ 
U_{22}P_2^{(2)}\otimes P_2^{(1)}
\end{eqnarray}
 The state of the system can then be written as- $|x\rangle=|m_1\rangle\otimes|m_2\rangle=|m_1m_2\rangle$. $|x\rangle$ represents the tensor product of the state of each site given by $m_i$ where $i$ is the site indexwith $m_{i} \in \{0,1\}$ . The partition function for the two spin system then is-
\begin{eqnarray}
Z&=&Tr(e^{-\beta H})=\!\!\sum_{x}\langle x|e^{-\beta H}|x\rangle=\!\!\sum_{m_1,m_2=0,1}\langle m_1m_2|(1-\beta H+\frac{(\beta H)^2}{2!}...)|m_1m_2\rangle\nonumber\\
&=& 1-\sum_{m_1,m_2=0,1}\beta\langle m_1m_2|H|m_1m_2\rangle+\!\!\sum_{m_1,m_2=0,1}\frac{\beta}{2!}\langle m_1m_2|H^2|m_1m_2\rangle +....
\end{eqnarray} 
Since $P_m^{1}=|m_1\rangle\langle m_1|$ and $P_m^{2}=|m_2\rangle\langle m_2|$, we have $H|m_1 m_2\rangle = U_{m_1m_2}|m_1m_2\rangle+U_{m_2m_1}|m_2m_1\rangle$. Thus,
\begin{eqnarray}
\langle m_1m_2|H|m_1m_2\rangle&=& U_{m_1m_2}+U_{m_2m_1},\\
\mbox{and  } \langle m_1m_2|H^2|m_1m_2\rangle &=&(U_{m_1m_2}+U_{m_2m_1})\langle m_1m_2|H|m_1m_2\rangle\nonumber\\&=&(U_{m_1m_2}+U_{m_2m_1})^2
\end{eqnarray}
Therefore, from Eq.~(33) we have-
\begin{eqnarray}
Z&=&\!\!\sum_{m_1,m_2=0,1}[1-\beta(U_{m_1m_2}+U_{m_2m_1})+\frac{\beta^2}{2!}(U_{m_1m_2}+U_{m_2m_1})^2+....],\nonumber\\
\mbox{or, }Z&=&\!\!\!\sum_{m_1,m_2=0,1}e^{-\beta(U_{m_1m_2}+U_{m_2m_1})}= e^{-2\beta U_{11}}+2e^{-\beta (U_{12}+ U_{21})}+e^{-2\beta U_{22}}.
\end{eqnarray}
In Ref.~[5], the condition: $U_{12}+U_{21}=U_{11}+U_{22}$ leads to-
\begin{eqnarray}
Z=(e^{-\beta U_{11}}+e^{-\beta U_{22}})^2.
\end{eqnarray}
Now for the public goods game the payoff matrix is-
\begin{equation}
U=\left(\begin{array}{c|cc} & s_1 & s_2 \\\hline s_1 & 2r,2r & r-\frac{c}{2},r+\frac{c}{2} \\  s_2 & r+\frac{c}{2},r-\frac{c}{2} & 0,0\end{array}\right)\nonumber
\end{equation} 
wherein $U_{11}=2r, U_{12}=r-\frac{c}{2}, U_{21}=r+\frac{c}{2}, U_{22}=0$ and because it's a symmetric game the  condition $U_{11}+U_{22}=U_{12}+U_{21}$ is satisfied. Thus, we have
\begin{eqnarray}
Z=(e^{-\beta U_{11}}+e^{-\beta U_{22}})^2=Tr(E^2)=(e^{-\beta w}+e^{-\beta x})^2\nonumber
\end{eqnarray}
From section 3 of our main manuscript, the energies in Ising model are equivalent to payoffs of game theory. Thus, we try to find the average payoff (or, the average energy). However, while energies are minimized to get the equilibrium (point of lowest energy or ground state) the payoffs in a game are maximized. The average payoff using Adami and Hintze's approach then is-
\begin{eqnarray}
\langle E\rangle=-\frac{\partial \ln Z}{\partial \beta}=-2\frac{\partial \ln((e^{-\beta U_{22}}+e^{-\beta U_{11}}))}{\partial \beta}=2\frac{(U_{22} e^{-\beta U_{22}}+U_{11} e^{-\beta U_{11}})}{(e^{-\beta U_{22}}+e^{-\beta U_{11}})}.
\end{eqnarray}
Now, in the limit $\beta\rightarrow\infty$, both players defect, i.e., there is no randomization of strategies. In this limit, we should get back the results for 2 player case. Thus imposing the limit $\beta\rightarrow\infty$, the payoff for each player (dividing average $\langle E\rangle$ by $2$) becomes-
\begin{eqnarray}\label{eq9}
\lim_{\beta\rightarrow\infty}\langle E\rangle=\lim_{\beta\rightarrow\infty}\frac{(U_{22} e^{-\beta U_{22}}+U_{11} e^{-\beta U_{11}})}{(e^{-\beta U_{22}}+e^{-\beta U_{11}})}=\lim_{\beta\rightarrow\infty}\frac{ 2r e^{-\beta 2r}}{1+e^{-\beta 2r}}
\end{eqnarray}
Let's take two cases and see whether the average payoff given via Eq.~(\ref{eq9}) gives  the correct average payoff- 
\begin{enumerate}
\item For $r>\frac{c}{2}$ in which case the Nash equilibrium is the strategy $(provide,provide)$. However, from Eq.~(\ref{eq9}), we get
\begin{eqnarray}
\lim_{\beta\rightarrow\infty}\langle E\rangle=\lim_{\beta\rightarrow\infty}\frac{ 2r e^{-\beta 2r}}{1+e^{-\beta 2r}}=0,
\end{eqnarray}
which is the payoff of the strategy $(freeride,freeride)$. In this case we arrive at a wrong conclusion.
\item  For $r<\frac{c}{2}$ the Nash equilibrium is the strategy $(freeride,freeride)$. From Eq.~(\ref{eq9}), we get-
\begin{eqnarray}
\lim_{\beta\rightarrow\infty}\langle E\rangle=\lim_{\beta\rightarrow\infty}\frac{ 2r e^{-\beta 2r}}{1+e^{-\beta 2r}}=0,
\end{eqnarray}
\end{enumerate}
which is the payoff of the strategy $(freeride,freeride)$ which are the correct payoffs for this case. Thus, we have shown that even in the two player case, the approach of Ref.~\cite{1} gives incorrect average payoffs for the case $r>\frac{c}{2}$ while it gives correct payoffs for $r<\frac{c}{2}$. Now let's analyze the two player case using our approach.\\
{\bf {Using our approach}}\\
For two spin case the Hamiltonian from Eq.~(1) of this manuscript is,
\begin{equation}
H=-J(\sigma_1\sigma_{2}+\sigma_2\sigma_{1})-h/2(\sigma_1+\sigma_2),
\end{equation}
herein $\sigma_1$ represents the spin at first site which can be either in one of two states: up or down. Similarly, $\sigma_{2}$ represents the spin at second site. The partition function of this two site system is-
\begin{equation}
Z=\sum_{\sigma_1}\sum_{\sigma_2}e^{\beta(J(\sigma_1\sigma_{2}+\sigma_2\sigma_{1})+h/2(\sigma_1+\sigma_2))}=e^{\beta(2J+h)}+2e^{-\beta(2J)}+e^{\beta(2J-h)}
\end{equation}
 For the public goods game, see Eq.~(\ref{trans-pg-payoff}), we get $J=0$ and $h=\frac{2r-c}{4}$. Thus, we have $Z=e^{\beta h}+2+e^{-\beta h}$.  Since the energy in Ising model is the equivalent of the payoffs of game theory but with the caveat that instead of minimizing the energies the payoffs are maximized. Thus, the average payoff for 2-player case is-
\begin{eqnarray}\label{eq15}
\langle E\rangle=-\frac{\partial \ln Z}{\partial \beta}=-\frac{\partial \ln(e^{\beta h}+2+e^{-\beta h}))}{\partial \beta}=\frac{-h e^{\beta h}+h e^{-\beta h}}{e^{\beta h}+2+e^{-\beta h}}
\end{eqnarray}
However, as was mentioned in section 3 of the main manuscript, the payoffs should be taken as $-E$ as for game we intend to maximize the payoffs (whereas, in solution to Ising model problem our aim is to minimize the energies, so as to determine the ground state). Now, let's consider the two cases and see whether our approach gives the correct average payoff.
\begin{enumerate}
\item For $r>\frac{c}{2}$ we have $h>0$ and the Nash equilibrium for the public goods game is the strategy $(Provide,Provide)$. Using our approach, from Eq.~(\ref{eq15}), we get-
\begin{eqnarray}
\lim_{\beta\rightarrow\infty}-\langle E\rangle=\lim_{\beta\rightarrow\infty}\frac{he^{\beta h}-he^{-\beta h}}{e^{\beta h }+2+e^{-\beta h }}=h=\frac{2r-c}{4}
\end{eqnarray}
which is the payoff of the strategy $(provide,provide)$ of the transformed payoff matrix, see Eq.~(\ref{trans-pg-payoff}) of the main manuscript.
\item  For $r<\frac{c}{2}$ we have $h<0$  and the Nash equilibrium for the public goods game is the strategy $(free ride, free ride)$. Using our approach, from Eq.~(\ref{eq15}), we get-
\begin{eqnarray}
\lim_{\beta\rightarrow\infty}-\langle E\rangle=\lim_{\beta\rightarrow\infty}\frac{he^{\beta h}-he^{-\beta h}}{e^{\beta h }+2+e^{-\beta h }}=-h=-\frac{2r-c}{4},
\end{eqnarray}
which is the payoff of the strategy $(free ride,free ride)$ of the transformed  payoff matrix, see Eq.~(\ref{trans-pg-payoff}) of the main manuscript.
\end{enumerate}
Thus, our approach gives the correct payoffs corresponding to the  Nash equilibrium strategies.\\
Ref.~[5]: C. Adami and A. Hintze, Thermodynamics of Evolutionary Games, Phys. Rev. E 97, 062136 (2018).\\
An essay by P Ralegankar, Understanding Emergence of Cooperation using tools from Thermodynamics, available at:\\ http://guava.physics.uiuc.edu/~nigel/courses/569/Essays$\_$Spring2018/Files/Ralegankar.pdf 
\\
also comes to similar conclusions.
\subsection{Invariance of Nash equilibrium for transformed payoff matrix, see Eq.~(\ref{eq12})}
The payoff matrix for two player public goods game can be written as-
\begin{equation}\label{eqa1}
U=\left(\begin{array}{c|cc} & provide & free ride \\\hline provide & 2r,2r & r-\frac{c}{2},r+\frac{c}{2} \\  free ride & r+\frac{c}{2},r-\frac{c}{2} & 0,0\end{array}\right).
\end{equation} 
Transforming the elements of the payoff matrix by adding a factor $\lambda$ to column 1 and $\mu$ to column 2 for row player's payoffs and adding a factor $\lambda'$ to row 1 and $\mu'$ to row 2 for column player's payoffs   we get:
\begin{equation}\label{eqa2}
U=\left(\begin{array}{c|cc} & provide & free ride \\\hline provide & 2r+\lambda,2r+\lambda' & r-\frac{c}{2}+\mu, r+\frac{c}{2}+\lambda'\\  free ride & r+\frac{c}{2}+\lambda,r-\frac{c}{2}+\mu' & 0+\mu, 0+\mu'\end{array}\right).
\end{equation} 
Making the transformations $\lambda=\lambda'=-\frac{6r+c}{4}$  and $\mu=\mu'=-\frac{2r-c}{4}$ such that the mapping of the $2$ players Public goods game to $2$ spin Ising energy matrix is exact as shown for general case of Eq.~(7), we get-
\begin{equation}\label{eqax}
U=\left(\begin{array}{c|cc} & provide & free\ ride \\\hline provide & \frac{2r-c}{4}, \frac{2r-c}{4}& \frac{2r-c}{4},-\frac{2r-c}{4} \\  free\ ride & -\frac{2r-c}{4},\frac{2r-c}{4} & -\frac{2r-c}{4},-\frac{2r-c}{4}\end{array}\right)
\end{equation} 
Our claim is that the Nash equilibrium in case of Eqs.~(\ref{eqa1}),~(\ref{eqax}) is identical. Below a simple proof of this claim using fixed point analysis is provided. A fixed point is a point in coordinate space which maps a function to the coordinate. In a two dimensional coordinate space, a fixed point of a function $f(x,y)$ is mathematically defined as $(x,y)$ such that\cite{11}- 
\begin{eqnarray}\label{eqa6}
f(x,y)=(x,y).
\end{eqnarray}
From Brouwer's fixed point theorem it is known that {\it a 2D triangle $\Delta_2$ has a fixed point property}. This implies that any function which defines all the points inside a 2D triangle has a fixed point (for a detailed proof of this theorem refer to \cite{11}). Also the probabilities for choosing a strategy, represents points inside a square of side length $1$. Thus
$S_{2,2}=(x,y)$ with $0<x<1$ and $0<y<1$, where $x$ represents the probability of choosing a strategy by row player and $y$ represents the probability of choosing a strategy by column player. It can be shown that a triangle and a square are topologically equivalent \cite{11}, and this implies that if a triangle has a fixed point property, so does a square. A function can be defined such that it represents all the points inside the square. To define such a function\cite{11}, lets consider a vector with coordinates ($u_1$, $u_2$) and another vector with coordinate ($v_1$, $v_2$) which are given as follows-
\begin{equation}\label{eqa9}
\left(\begin{array}{c} u_1\\u_2\end{array}\right)=A \left(\begin{array}{c} y\\1-y\end{array}\right),
\end{equation}
and
\begin{equation}\label{eqa10}
\left(\begin{array}{cc} v_1 & v_2\end{array}\right)= \left(\begin{array}{cc} x & 1-x\end{array}\right)B,
\end{equation}
where $x$ and $y$ are the probabilities to choose a particular strategy. $A$ and $B$ denote the respective payoff matrix for row player and column player. Using this, the fixed point function from Eq.~(\ref{eqa6}) is given by
\begin{equation}\label{eq44}
f(x,y)=\left(\frac{x+(u_1-u_2)^+}{1+|u_1-u_2|},\frac{y+(v_1-v_2)^+}{1+|v_1-v_2|}\right),
\end{equation}
where $(u_1-u_2)^+=\frac{u_1-u_2+|u_1-u_2|}{2}$ and $(v_1-v_2)^+=\frac{v_1-v_2+|v_1-v_2|}{2}$.
We determine $u_1$, $u_2$, $v_1$ and $v_2$ for the  payoff matrix as in Eq.~(\ref{eqa1}) and then the transformed one as in Eq.~(\ref{eqa2}). For the payoff matrix Eq.~(\ref{eqa1}) we get the coordinates ($u_i$ and $v_i$ for i=1,2) as
\begin{eqnarray}\label{eqa3}
u_1=2r y+(r-\frac{c}{2}) (1-y)\nonumber\\
u_2=(r+\frac{c}{2}) y+0 (1-y)\nonumber\\
v_1=2r x+ (r-\frac{c}{2}) (1-x)\nonumber\\
v_2=(r+\frac{c}{2}) x+0 (1-x).
\end{eqnarray}
Now for the transformed payoff matrix as in  Eq.~(\ref{eqa2}) the fixed point function is given by
\begin{equation}\label{eqa5}
f^t(x,y)=\left(\frac{x+(u_1^t-u_2^t)^+}{1+|u_1^t-u_2^t|},\frac{y+(v_1^t-v_2^t)^+}{1+|v_1^t-v_2^t|}\right).
\end{equation}
Again we determine the coordinates ($u_i^t$ and $v_i^t$ for i=1,2), as
\begin{eqnarray}\label{eqa4}
u_1^t=\frac{2r-c}{4} y+\frac{2r-c}{4} (1-y)\nonumber\\
u_2^t=-\frac{2r-c}{4} y -\frac{2r-c}{4}(1-y)\nonumber\\
v_1^t=\frac{2r-c}{4}x+\frac{2r-c}{4}(1-x)\nonumber\\
v_2^t=-\frac{2r-c}{4} x-\frac{2r-c}{4} (1-x).
\end{eqnarray}
From Eq.~(\ref{eqa3}) and Eq.~(\ref{eqa4}), $u_1-u_2=u_1^t-u_2^t=r-\frac{c}{2}$ and $v_1-v_2=v_1^t-v_2^t=r-\frac{c}{2}$. Thus, $f^t(x,y)=f(x,y)$ which implies that the Nash equilibrium remains unchanged under the transformations as described before in Eqs.~(\ref{eqa1},\ref{eqa2}). 
\section{Acknowledgments} This work was supported by the grants- 1. ``Non-local correlations in nanoscale systems: Role of decoherence, interactions, disorder and pairing symmetry'' from SCIENCE \& ENGINEERING RESEARCH BOARD, New Delhi, Government of India, Grant No.  EMR/20l5/001836, Principal Investigator: Dr. Colin Benjamin, National Institute of Science Education and Research, Bhubaneswar, India,  and 2. ``Nash equilibrium versus Pareto optimality in N-Player games'', SERB MATRICS Grant No. MTR/2018/000070, Principal Investigator: Dr. Colin Benjamin, National Institute of Science Education and Research, Bhubaneswar, India.


\begin{thebibliography}{99}
\bibitem{5}Game Theory in Action: An Introduction to Classical and Evolutionary Models, S. Schecter and H. Gintis, Princeton University Press (2016).
\bibitem{11}S. B. Ale, J. S. Brown, A. T. Sullivan,  Evolution of Cooperation: Combining Kin Selection and Reciprocal Altruism into Matrix Games with Social Dilemmas, PLoS ONE 8(5): e63761 (2013).
\bibitem{12}M. A. Nowak, A. Sasaki, C. Taylor, and D. Fudenberg, Emergence of cooperation
and evolutionary stability in finite populations, Nature, 428, 646 (2004).
\bibitem{19}G. Szabo and I. Borsos, Evolutionary potential games on lattices, Physics Reports, 624, 1 (2016).
\bibitem{1} C. Adami and A. Hintze, Thermodynamics of Evolutionary Games, Phys. Rev. E 97, 062136 (2018).
\bibitem{8}S. Sarkar and C. Benjamin, Emergence of Cooperation in the thermodynamic limit, arXiv:1803.10083.
\bibitem{2}S. Galam and B. Walliser, Ising model versus normal form game, Physica A, 389, 481 (2010).
\bibitem{6} S. Salinas, Introduction to Statistical Physics, 1st Edition, Springer (2000).
\bibitem{7}C. Hauert and G. Szabo, Prisoner's dilemma and public goods games in different geometries: Compulsory versus voluntary interactions, Complexity, 8, 31 (2003).
\bibitem{11} Game Theory: A Playful Introduction, M. DeVos and D. A. Kent, American Mathematical Society (2016)
\end{thebibliography}
\end{document}